\documentclass[manuscript]{aastex}
\usepackage{multirow}
\usepackage{amsmath}
\usepackage{color}

\slugcomment{}

\shorttitle{Detailed molecular observations toward the Double Helix Nebula}
\shortauthors{K. Torii et al.}

\begin{document}


\title{Detailed molecular observations toward the Double Helix Nebula}


\author{K. Torii\altaffilmark{1, 2}, R. Enokiya\altaffilmark{1}, M.R.Morris\altaffilmark{3}, K. Hasegawa\altaffilmark{1}, N. Kudo\altaffilmark{1} and Y. Fukui\altaffilmark{1}}
\affil{$^1$Department of Physics and Astrophysics, Nagoya University, Chikusa-ku, Nagoya, Aichi, 464-8601, Japan}
\affil{$^2$Sub-Department of Astrophysics, University of Oxford, Denys Wilkinson Building, Keble Road, Oxford OX1 3RH, UK}
\affil{$^3$Department of Physics and Astronomy, University of California, Los Angeles, California 90095-1547, USA.}

\email{torii@a.phys.nagoya-u.ac.jp}

\begin{abstract}

The Double Helix Nebula (DHN), located 100 pc above Sgr A* in the Galactic center (GC), is a unique structure whose morphology suggests it is a magnetic feature \citep{mor2006}. 
Recent molecular observations toward the DHN by \citet{eno2013} revealed two candidate molecular counterparts of the DHN at radial velocities of $-35$ km s$^{-1}$ and 0 km s$^{-1}$ and discussed the model in which the DHN has its origin at the circumnuclear disk in the GC.
In this paper, new CO observations toward the DHN using the CSO and Mopra telescopes are presented. 
The higher-resolution observations of $\sim$1 pc scale reveal the detailed distributions and kinematics of the two CO counterparts (the 0 km s$^{-1}$ and $-35$ km s$^{-1}$ features) and provide new information on their physical conditions. 
As a result, we find that the 0 km s$^{-1}$ feature with a mass of $3.3\times10^4$ M$_\odot$ coincides with the infrared emission of the DHN, indicating clear association with the DHN.
The association of the $-35$ km s$^{-1}$ feature, with a mass of $0.8\times10^4$ M$_\odot$, is less clear compared with the 0 km s$^{-1}$ feature, but the complementary distribution between the molecular gas and the DHN and velocity variation along the DHN support its association with the DHN. 
The two molecular features are highly excited, as shown by the relatively high CO $J$=2--1/$J$=1--0 intensity ratios of $\sim$1.0, and have kinetic temperatures of $\sim$30 K, consistent with the typical molecular clouds in the GC. 

\end{abstract}


\section{Introduction}
The Galactic center (GC) has many outstanding structures that are not seen in the outer part of the Galaxy. 
In particular, several lines of evidence indicate that the magnetic field plays an important role in this region. 
A strong magnetic field of 50 $\mu$G has been suggested as an averaged figure within the central 400 pc region by an analysis of the non-thermal radio spectrum \citep{cro2010}, and many unique astrophysical structures related to the magnetic field have been discovered so far. 
Many linear, non-thermal radio filaments are present in the Radio Arc \citep{yus1984}, in which highly ordered magnetic field lines distributed vertically to the Galactic plane are traced by their radio synchrotron emission \citep[e.g.,][]{lan1999,lar2000,yus2004}.  
Their origin is still elusive, though numerous ideas have been advanced \citep{mor1996b}. 
On a larger scale, \citet{fuk2006} discovered two giant molecular loops $\sim$700 pc away from the center with a height of $\sim$200 pc, and they propose that they are a result of magnetic buoyancy driven by the Parker instability. 
Follow-up studies reveal that the footpoints of the loops have highly turbulent molecular clumps with velocity dispersions of $\sim$50 km s$^{-1}$, and magnetic reconnection has been discussed as the origin of these clumps \citep{tor2010a,tor2010b,kud2011}.

The Double Helix Nebula (hereafter DHN) was discovered $\sim$100 pc above the Galactic center by \citet{mor2006} using infrared observations with Spitzer (Figure \ref{dhn}). 
It has an apparent helical morphology that can be seen in dust emission, implying that it is organized by a magnetic field.  
\citet{mor2006} suggest that the DHN was created by torsional Alfv\'en waves emitted from the circum-nuclear disk (CND) which surrounds the supermassive black hole, Sgr A*. 
On the other hand, \citet{tsu2010} use radio polarization measurements to hypothesize that the DHN is an extension of the polarized northern lobe of the magnetic Radio Arc. 
An interesting clue that supports the magnetic nature of the DHN is the presence of non-thermal radio emission distributed along the western rim of the DHN \citep{law2008}.

Most recently, \citet{eno2013} present the observations of a $4^\circ \times 2^\circ$ area of the GC in the $^{12}$CO($J$=2--1) transition obtained using the NANTEN2 4m telescope with a beam size of 100$''$, finding that two molecular features at radial velocities of $\sim-$35 km s$^{-1}$ and 0 km s$^{-1}$ (hereafter the $-$35 km s$^{-1}$  and 0 km s$^{-1}$ features) coincide with the DHN. 
They also find that these features are located at the tops of molecular ridges elongated vertically to the Galactic plane, having lengths of $\sim$150 pc at the GC distance. 
Indeed, they estimate the distance of the ridges as 8$\pm$2 kpc, which is consistent with the distance to the GC, by carrying out an analysis of the $K$-band stellar extinction. 
It therefore seems quite likely that the DHN and its molecular counterparts have their origin at the GC.
However, a spatial resolution of 100$''$, which corresponds to $\sim$4 pc at the GC, is much coarser than the typical size of the helical filaments of the DHN, $\sim$1--2 pc, and detailed comparisons of molecular emission with the 24 $\mu$m Spitzer image have not yet been possible.

In this study, we present results of new molecular observations toward the DHN using the CSO and Mopra telescopes. 
The improved spatial resolutions of $\sim$33$''$ ($\sim$1.3 pc) enable a more detailed description of the $-35$ km s$^{-1}$ and 0 km s$^{-1}$ features and help clarify whether they are physically associated with the DHN. 
This paper is organized as follows; Section 2 summarizes the observations and Section 3 the results. 
The discussion is given in Section 4 and a summary in Section 5. 
In this study, we adopt a GC distance of 8.0 kpc.

\section{Observations}
\begin{figure}
\epsscale{.40}
\plotone{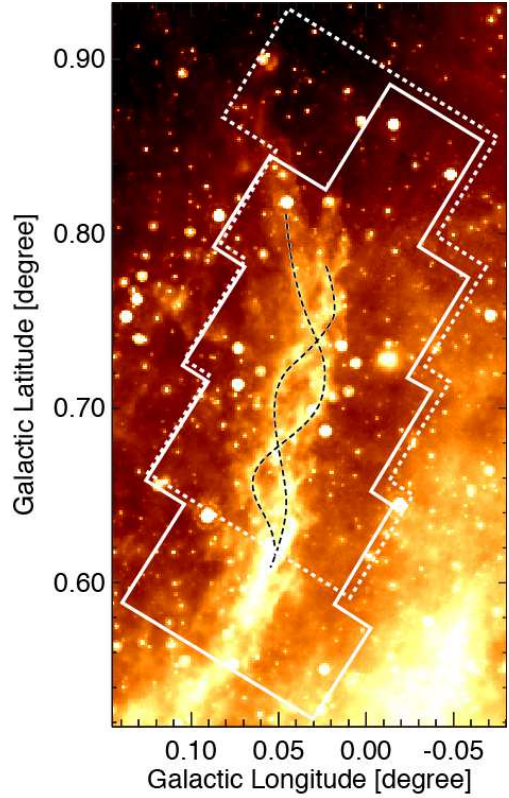}
\caption{The Spitzer 24 $\mu$m flux density distribution toward the DHN \citep{mor2006}. The approximate ridges of the two strands of the DHN are indicated by dashed lines. The observed areas of the CSO and Mopra observations are depicted by solid and dotted lines, respectively.\label{dhn}}
\end{figure}

Observations of the $^{12}$CO($J$=2--1) and $^{13}$CO($J$=2--1) transitions were carried out with the 10.4m telescope of the Caltech Submillimeter Observatory (CSO) in June 2011. 
The half-power beamwidth (HPBW) of the telescope was 33$''$ at 230 GHz, which corresponds to 1.4 pc at the distance of the GC. 
The 230 GHz SIS receiver provided a typical single-sideband (SSB) system temperature, $T_{\mathrm{sys}}$, of $\sim$400--500 K at 230 GHz and enabled us to observe the $^{12}$CO($J$=2--1) and $^{13}$CO($J$=2--1) transitions simultaneously. 
The spectrometer was a Fast Fourier Transform Spectrometer (FFTS) with 8192 channels, providing a velocity coverage of $\sim$1300 km s$ ^{-1}$ with a velocity resolution of 0.16 km s$^{-1}$ at 230 GHz.   
We observed the nine 4$'$ $\times$ 4$'$ regions shown in Figure \ref{dhn} with the on-the-fly (OTF) mode. 
The output grid of the observations is spaced by 15$''$. We smoothed the velocity resolution and spatial resolution to 0.79 km s$^{-1}$ and 45$''$, respectively, to achieve a better noise level. 
The pointing accuracy was checked and adjusted every hour to be better than 5$''$ using observations of SiO masers. 
The obtained spectra were converted into a $T_{\mathrm{mb}}$ scale with a beam efficiency of 0.698. 
In the final maps, we obtained r.m.s. noise fluctuations of $\sim$0.08 K per channel. 

Observations of the $^{12}$CO($J$=1--0) and $^{13}$CO($J$=1--0) transitions were carried out using the 22m ATNF Mopra mm telescope in Australia in October 2011, which provided a HPBW of 33$''$ at 110 GHz, the same as that of the $J$=2--1 lines measured at the CSO. 
The OTF mode was again used toward the eight 4$'$ $\times$ 4$'$ regions shown in Figure \ref{dhn}. 
The typical SSB system noise temperature was 500 K. 
The Mopra backend system ``MOPS'' provides 4096 channels across 137.5 MHz in each of the two orthogonal polarizations. 
The effective velocity resolution was 0.088 km s$^{-1}$ and the velocity coverage was 360 km s$^{-1}$ at 115 GHz. 
The pointing accuracy was checked every hour to be better than 7$''$ using observations of 86 GHz SiO masers. 
The spectra were gridded to a 15$''$ spacing, and smoothed to a 45$''$ beam size. 
We observed Orion-KL (RA, Dec.)=($5^{\rm h}35^{\rm m}14\fs5$, $-5\degr22\arcmin29\farcs6$) for the absolute intensity calibration and carried out comparisons with the results of \citet{lad2005}. 
The ``extended beam efficiency'' in \citet{lad2005} was estimated to be 0.38 by dividing the observed peak antenna temperature of Orion-KL by 100 K, which is the corrected absolute intensity shown in Figure 6 of \citet{lad2005}. 
After calibrating the spectra, we finally smoothed the channels in velocity to a resolution of 0.86 km s$^{-1}$. 
The typical r.m.s noise levels achieved in the $^{12}$CO($J$=1--0) and $^{13}$CO($J$=1--0) spectra are 0.14 K and 0.07 K, respectively.

The reference position of the OTF scans used in the CSO and Mopra observations is $(l, b)=(359\fdg334, 1\fdg800)$, which is chosen to have no significant emission over a velocity range of $\pm$100 km s$^{-1}$ at an r.m.s noise level of $\sim$0.05 K per channel in both transitions.

\section{Results}
\subsection{Spatial and velocity distribution of the molecular gas}
\begin{figure}
\epsscale{.90}
\plotone{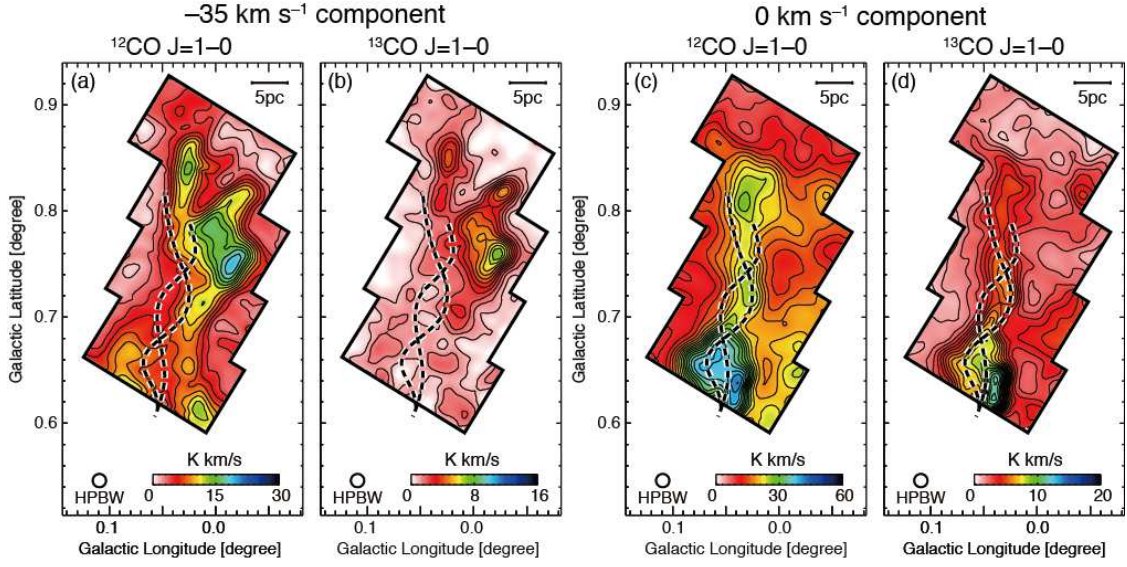}
\caption{Distributions of the two CO features in $J$=1--0 emission. The $-35$ km s$^{-1}$ feature is shown in panels (a) and (b), and the 0 km s$^{-1}$ feature in panels (c) and (d). Panels (a) and (c) show the emission from $^{12}$CO($J$=1--0), and panels (b) and (d) show the $^{13}$CO($J$=1--0) emission. The contour levels of each panel are as follows; (a) minimum: 2 K km s$^{-1}$, step: 1.5  K km s$^{-1}$. (b) minimum: 0.8 K km s$^{-1}$, step: 0.6 K km s$^{-1}$. (c) minimum: 2 K km s$^{-1}$, step: 2 K km s$^{-1}$. (d) minimum: 0.8 K km s$^{-1}$, step: 0.6 K km s$^{-1}$.
\label{lb}}
\end{figure}

In Figure \ref{lb} we present the integrated intensity distributions of the $-35$ km s$^{-1}$ feature and the 0 km s$^{-1}$ feature in $^{12}$CO($J$=1--0) and $^{13}$CO($J$=1--0).
The $-35$ km s$^{-1}$ feature shown in Figures \ref{lb}(a) and \ref{lb}(b) is distributed between $-36$ and $-30$ km s$^{-1}$, widely covering the whole DHN.
It extends to $b\sim0\fdg86$, beyond the top of the DHN where the 24 $\mu$m emission becomes too faint to discern any possible continuation of the helical pattern.
The $-35$ km s$^{-1}$ CO gas shows the strongest emission just to the west and north of the DHN and is spatially complementary to the western edge of the DHN, suggesting a possible association.
In Figures \ref{lb}(c) and \ref{lb}(d), the 0 km s$^{-1}$ feature shows a remarkably good coincidence with the DHN. 
It has brighter CO emission and traces the helical distribution of the DHN.
The width of the 0 km s$^{-1}$ feature is $\sim$4 pc, comparable to that of the DHN. 
The 0 km s$^{-1}$ feature is also extended to $b\sim0\fdg86$, where the infrared emission also fades out.
The  excellent spatial coincidence between the CO and infrared emission provides robust evidence that the 0 km s$^{-1}$ feature is physically associated with the DHN, while association of the $-35$ km s$^{-1}$ feature is less clear.

In Appendix A, we show in Figure \ref{lb2-1} the integrated intensity of the CO($J$=2--1) emission, and in Figures \ref{channel_all1} -- \ref{channel_0}, the velocity channel maps of $^{12}$CO($J$=1--0) and $^{12}$CO($J$=2--1) emission.
Another CO feature around 8 to 15 km s$^{-1}$ (hereafter the 10 km s$^{-1}$ feature) overlaps the DHN, as seen in Figure \ref{channel_all1}, but the gas distribution does not coincide with the DHN, thus it seems not to be associated. 
This feature is also seen in the large scale view of \citet{eno2013}, and they conclude that a relationship with the DHN is unlikely.

\begin{figure}
\epsscale{.5}
\plotone{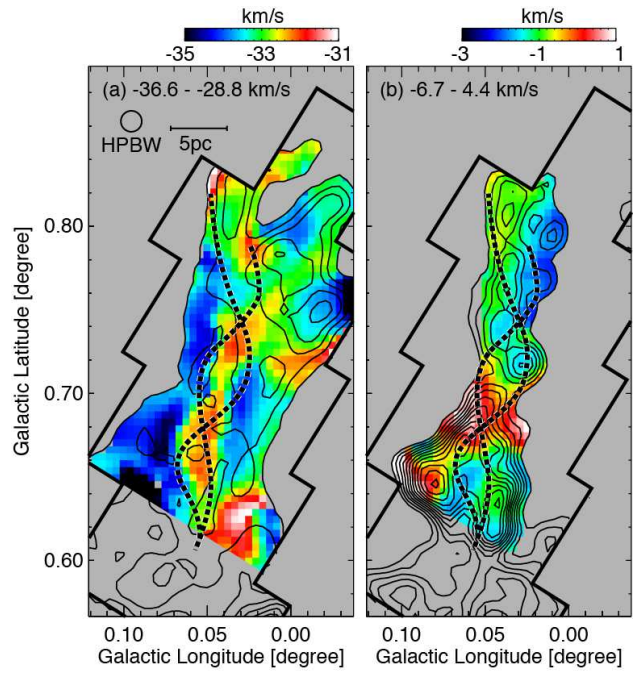}
\caption{First moment maps of the $-35$ km s$^{-1}$ (left) and 0 km s$^{-1}$ features (right). Contours show the integrated $^{12}$CO($J$=2--1) emission shown in Figure \ref{lb}(b)--(c).\label{peakv}}
\end{figure}

Figures \ref{peakv}(a) and \ref{peakv}(b) show the distributions of the first moment maps of the spectra in the $-$35 km s$^{-1}$ and 0 km s$^{-1}$ features, respectively. 
Here $^{12}$CO($J$=2--1) is used, which has less complicated profiles than $^{12}$CO($J$=1--0).
In the $-35$ km s$^{-1}$ feature, a velocity jump up to 2 km s$^{-1}$ which coincides spatially with the DHN is seen at $(l, b)\sim(0.05^\circ, 0.65$--$0.75)$. 
As can be seen in Figure \ref{channel_m35} of Appendix A, the $-35$ km s$^{-1}$ feature consists of three filamentary structures that are distributed almost parallel to the DHN; two of them occur around $-35$ km s$^{-1}$ and the third one, which coincides with the DHN, is found near $-31$ km s$^{-1}$.
The apparent velocity jump seen in Figure \ref{peakv}(a) is due to the velocity difference between these filamentary structures.
In Figure \ref{peakv}(b), around the top of the DHN ($b\sim0\fdg75$--$0\fdg80$), the 0 km s$^{-1}$ feature shows an east-west velocity gradient of 1 km s$^{-1}$ pc$^{-1}$.
At the lower-latitude part of the 0 km s$^{-1}$ feature, the velocity distribution is much more complicated.
We find a velocity jump of 2 km s$^{-1}$ around the intersection of the DHN strands at $b\sim0\fdg68$.

\begin{figure}
\epsscale{.45}
\plotone{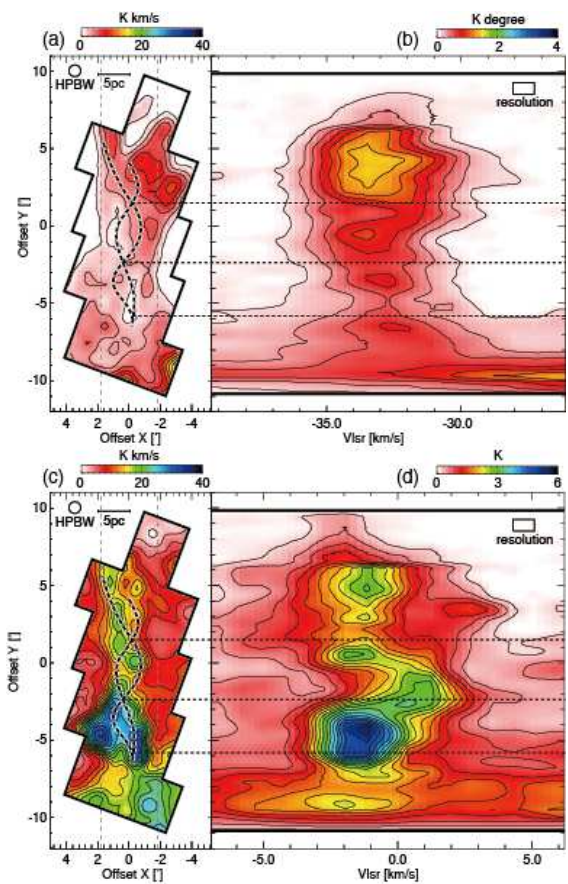}
\caption{
(a, c) Integrated intensity of $^{12}$CO($J$=2--1) emission in the $-35$ km s$^{-1}$ and 0 km s$^{-1}$ features, respectively, oriented so that the long axis of the DHN is vertical. (b, d) Velocity-position (Y-axis) distributions of $^{12}$CO($J$=2--1) integrated between the vertical lines in panels (a) and (c). The positions of the intersections of the DHN strands are shown by horizontal thin dotted lines. 
\label{lb+bv_0+m35}}
\end{figure}

To show the velocity distributions of the $-35$ km s$^{-1}$ and 0 km s$^{-1}$ features in more detail, we define a new coordinate parallel to the DHN. 
A linear fit to the infrared emission of the DHN shows it to be tilted at a position angle of $-$11\fdg6 relative to Galactic north. 
We adopt the term ``DHN coordinate'' to refer to positions along this axis, with the origin taken to be $(l,b)\sim(0\fdg03, 0\fdg72)$.
Figures \ref{lb+bv_0+m35}(a) and \ref{lb+bv_0+m35}(c) show the $^{12}$CO($J$=2--1) distributions of the two velocity features in this DHN orientation, and Figures \ref{lb+bv_0+m35}(b) and \ref{lb+bv_0+m35}(d) show the velocity-position distributions. 
In Figure \ref{lb+bv_0+m35}(b) the top of the $-$35 km s$^{-1}$ feature shows a velocity width of $\sim$5 km s$^{-1}$, broader than the rest of the feature. 
Figure \ref{lb+bv_0+m35}(d) shows a sinuous velocity variation between $-5'$ and $0'$ along the Y-axis, which results largely from the velocity jump found at $b\sim0\fdg68$ in Figure \ref{peakv}(b).

\begin{figure}
\epsscale{.8}
\plotone{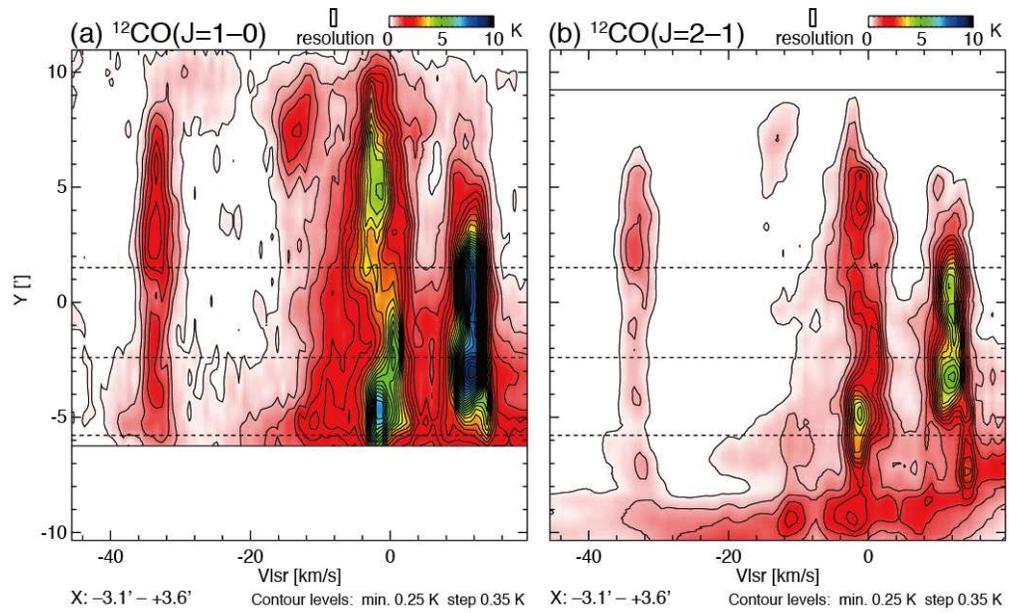}
\caption{$^{12}$CO($J$=1--0) (a) and $^{12}$CO($J$=2--1) (b) velocity-position (Y-axis) distributions of the $-35$ km s$^{-1}$ and 0 km s$^{-1}$ features. Horizontal dotted lines show the positions of the intersections of the DHN strands. \label{allbv}}
\end{figure}

Figure \ref{allbv} shows the velocity-position distributions of the $^{12}$CO($J$=1--0) and $^{12}$CO($J$=2--1) emission over a large velocity range, $-45$~--~$+20$ km s$^{-1}$ along the DHN coordinate. 
The 0 km s$^{-1}$ feature and the 10 km s$^{-1}$ feature are surrounded by diffuse molecular gas which is likely located in the foreground, and thus not necessarily associated with the DHN features.
Essentially no foreground emission is seen around the $-35$ km s$^{-1}$ feature, especially in the $^{12}$CO($J$=2--1) line. 
A CO component at the Y-axis $\simeq5'$~--~8$'$ and velocity of $-18$~--~$-10$ km s$^{-1}$ can be seen on the northwest side of the DHN in Figure \ref{channel_all1} of Appendix A.
There is no morphological indication that it is related to the DHN.
Diffuse CO emission around the Y-axis $\simeq8'$~--~11$'$ (Galactic latitude $\sim$ 0.9$^\circ$) and velocity of $-30$~--~$-22$ km s$^{-1}$ can be seen in Figure \ref{allbv}a; it might be spatially connected with the $-35$ km s$^{-1}$ feature as seen in Figure \ref{channel_all1} of Appendix A.
The broad velocity emission evident at the Y-axis $\simeq-10'$~--~$-8'$ lies at the northern edge of ``the connecting feature'' identified by \citet{eno2013}, which possibly links the $-35$ feature and the 0 km s$^{-1}$ features.

\subsection{Excitation conditions of the molecular gas}
Next we discuss the excitation condition of the molecular gas in the $-35$ km s$^{-1}$ and 0 km s$^{-1}$ features. 
Figure \ref{ratio_all} shows the distributions of the $^{12}$CO $J$=2--1/$J$=1--0 velocity-integrated intensity ratios for the $-30$ km s$^{-1}$, 0 km s$^{-1}$ and 10 km s$^{-1}$ features.
The individual velocity channel distributions of the line ratios for each of these features are presented in Figures \ref{ratiochannel_m35} -- \ref{ratiochannel_10} in Appendix A.
The $-35$, 0, and 10 km s$^{-1}$ features all show typical ratios of $\sim$0.8 and up to 1.1, whereas the foreground emission which is seen at velocities between $-7$ and $-4$ km s$^{-1}$ in Figure \ref{ratiochannel_0} shows lower ratios of less than 0.4.
Figure \ref{lvg_plot} shows curves of the ratio as a function of temperature $T_{\rm k}$ and density $n$(H$_2$) derived using LVG calculations (see next subsection for details).
The ratio 0.8 indicates excited condition of molecular gas with high temperature $> 20$~K and/or high density $>10^{3}$ cm$^{-3}$, while the ratio 0.4 indicates only low temperature ($\sim 10$ K) at low density 10$^2$ cm$^{-3}$.
A high excitation condition of the molecular gas is one of the major characteristics of the CMZ.
The high observed CO ratios therefore lend support to the placement of the candidate molecular counterparts of the DHN --- the 0 and $-$35 km s$^{-1}$ features --- near the GC, as also discussed in \citet{eno2013}.
The 10 km s$^{-1}$ feature in Figure \ref{ratio_all}(c) also shows high excitation conditions and thus is perhaps located in the GC, although, as discussed above, the gas distribution of the 10 km s$^{-1}$ feature does not coincide spatially with the infrared emission of the DHN, suggesting that it is an independent, unrelated feature. 
We also note here that the less excited foreground emission can reduce the CO intensities and thus can alter the intensity ratios in the DHN feature by absorption.
We present model calculations of the absorption in Appendix B using the LVG assumption, indicating that, although absorption can alter the ratios, ratios larger than 0.9 still indicate high excitation gas in the DHN. 

\subsection{Physical conditions of the molecular gas}
We here estimate the total molecular masses of the $-35$ km s$^{-1}$ feature and the 0 km s$^{-1}$ feature. 
We adopt the X-factor -- the conversion factor from the integrated intensity of $^{12}$CO($J$=1--0) to H$_2$ column density -- of 0.7$\times$10$^{20}$ (K km s$^{-1}$)$^{-1}$ cm$^{-2}$ \citep{tor2010b}, 
and derive the masses of the $-35$ km s$^{-1}$ and 0 km s$^{-1}$ features as $0.8\times10^4$ M$_\odot$ and $3.3\times10^4$ M$_\odot$, respectively, consistent with previous estimates using NANTEN2 \citep{eno2013}.
As mentioned in the previous subsection, absorption by foreground gas can differentially reduce the CO intensities.
Our model calculations shown in Appendix B suggest that these total molecular masses can be underestimated by 30--80 \%. 

We next discuss $T_{\rm k}$ and $n$(H$_2$) of the $-35$ km s$^{-1}$ and 0 km s$^{-1}$ features. 
We utilize an LVG analysis \citep[e.g.,][]{gol1974} to estimate these parameters. 
For this analysis we take intensity ratios of different $J$-levels of the CO emission. 
We use the $^{12}$CO($J$=2--1), $^{13}$CO($J$=1--0) and $^{13}$CO($J$=2--1) emission for the analysis and do not include the $^{12}$CO($J$=1--0) emission, since subthermally excited $^{12}$CO($J$=1--0) could selectively trace the diffuse envelopes of molecular features differently and more predominantly than the other three lines.
We apply the LVG analysis to the five regions A--E in the $-35$ km s$^{-1}$ and 0 km s$^{-1}$ features identified in Figure \ref{spec}.
Here we use the intensity ratios integrated over the velocity ranges indicated in the spectra in Figure \ref{spec} shown by dotted lines.

We adopt the abundance ratios of [$^{12}$CO]/[$^{13}$CO]=24 \citep{lan1990, riq2010} and the fractional CO abundance to be $X$(CO)=[$^{12}$CO]/[H$_2$]$=10^{-4}$ \citep[e.g.,][]{fre1982,leu1984}. 
We estimate velocity gradients of 0.4 km s$^{-1}$ pc$^{-1}$ by assuming the observed velocity variations of $\sim$2 km s$^{-1}$ and the width of the DHN of $\sim$5 pc, therefore giving $X$(CO)/($dv/dr$) of $2.5\times10^{-4}$ (km s$^{-1}$ pc$^{-1}$)$^{-1}$. 
Figure \ref{lvg} shows the results of the calculations, where the distributions of the two line ratios, $^{13}$CO($J$=2--1)/$^{13}$CO($J$=1--0) and $^{13}$CO($J$=2--1)/$^{12}$CO($J$=2--1), are presented, and details of the results are summarized in Table \ref{lvglist}. 
Here the errors are estimated with 1) 1$\sigma$ baseline fluctuations of the spectra, 2) 10\% relative calibration error between the Mopra observations and the CSO observations (which is only applied when comparing between $J$=2--1 and $J$=1--0 and is applied only to $^{13}$CO($J$=2--1)/$^{13}$CO($J$=1--0) in the present analysis), and 3) absorption rate of up to 50\% for the $^{12}$CO($J$=2--1) intensity due to the foreground cool gas (applied only to $^{13}$CO($J$=2--1)/$^{12}$CO($J$=2--1)).
Absorption of $^{13}$CO is ignored.
The $n$(H$_2$) and $T_{\rm k}$ ranges covered by the present analysis are $10^2$~--~$10^4$~cm$^{-3}$ and 7~--~200~K, respectively. 
As a result, we deduced that all five regions have $T_{\rm k}$ of 20~--~40 K at $n$(H$_2$) $\sim10^3$ cm$^{-3}$, which is significantly higher than the typical $T_{\rm k}$ of molecular gas in the Galactic disk, $\sim$10 K. Temperatures of $\sim$100~K are allowed for regions B and D. These two regions are located around the top of the DHN.

\section{Discussion}

Our detailed observations of the DHN with Mopra and the CSO indicate that both the $-35$ km s$^{1}$ and 0 km s$^{-1}$ features are associated with the DHN. 
The spatial distribution of the 0 km s$^{-1}$ feature shows remarkably good correspondence with the infrared emission of the DHN.
The association of the $-35$ km s$^{-1}$ feature is less clear compared with the 0 km s$^{-1}$ feature, but the complementary distribution between the molecular gas and the DHN and velocity variation along the DHN support the association. 
The high excitation condition shown by the $^{12}$CO $J$=2--1/$J$=1--0 ratio (Figure \ref{ratio_all}) and estimated high temperature of about 30 K (Figure \ref{lvg}) also support the placement of the two molecular features within the CMZ.
 
The two competing scenarios for the DHN are currently 1) a torsional Alfv\'en wave launched from the CND \citep{mor2006} and 2) an extension of the polarized northern lobe of the magnetic Radio Arc \citep{law2008,tsu2010}.
\citet{eno2013} find that the $-35$ and 0 km s$^{-1}$ features are continuations to higher latitude of molecular ridges that extend down to the Galactic plane; they also find that they are located in the GC.
The ridge at 0 km s$^{-1}$ appears to be pointing toward the CND rather than to the Radio Arc, although this does not provide an incontrovertible clue supporting the CND hypothesis.
The results presented here for the detailed distribution of molecular emission toward the DHN indicate that the both the $-35$ and 0 km s$^{-1}$ features are associated with the DHN; this was not conclusive in \citet{eno2013}. 
These two molecular features cannot be easily understood with either of the two present scenarios, so further investigations including theoretical studies are required.

The warm the warm temperature of the molecular counterparts of the DHN is a characteristic that helps associate these features with the clouds of the Central Molecular Zone.
The energy injection rate required to keep the high temperature is estimated to be $1\times10^{36}$ erg s$^{-1}$ \citep{gol1978}. 
Here we assume a cylinder with a length of 30 pc for each feature to roughly estimate its volume, the diameters of which are estimated to be 10 pc and 5 pc for the $-35$ km s$^{-1}$ feature and 0 km s$^{-1}$ feature, respectively, and we also assume a uniform density of 10$^{3}$ cm$^{-3}$ and uniform temperature of the molecular gas, 30 K.
It is reasonable to expect that the heating mechanism for the molecular gas is similar to that operating throughout the CMZ since 30~K is within a range of the typical temperatures of molecular gas in the CMZ, although the heating mechanisms in the CMZ are still under active discussion.

One possibility is heating by UV radiation from the large population of massive stars in the CMZ. 
We estimate the total infrared luminosity toward the DHN ($l=0\fdg00$--$0\fdg08$, $b=0\fdg60$--$0\fdg85$) as $8.5\times10^5$ L$_\odot$ = $3.3\times10^{39}$ erg s$^{-1}$ using the integrated fluxes of the four IRAS bands and the equation given in the Table 1 of \citet{san1996}, 
3.5 orders of magnitude larger than the cooling energy of the gas estimated above.
In the 24 $\mu$m image in Figure \ref{dhn}, the diffuse emission distributed around the DHN accounts for roughly about 50~\% of the flux density of the DHN.
Thus, if we consider the contribution of the diffuse emission, the total infrared luminosity is still much larger than the cooling energy.
Stellar heating is therefore a possible explanation of the observed warm gas.

Cosmic-rays are also a possible heating source for the molecular gas in the DHN.
The non-thermal radio emission which is used to probe the distribution of the cosmic-ray electrons shows a distribution extending to high latitude \citep{yus2013}, including the northern part of the polarized lobe \citep{tsu2010}. 

Another way to heat molecular gas is dissipative heating of the kinetic energy of the turbulent gas via ion-neutral friction and/or magnetic reconnection.
In the low latitude region of the CMZ close to the Galactic plane, highly turbulent molecular gas with a strong magnetic field possesses plenty of energy, but the velocity widths of the molecular features in the DHN are only about a few km s$^{-1}$, much smaller than the typical figures in the CMZ, making it less likely as the origin of the warm molecular gas in the DHN.
  
\section{Summary}
The present study is summarized as follows;

\begin{enumerate}
\item We have carried out new observations of $^{12}$CO($J$=1--0), $^{13}$CO($J$=1--0), $^{12}$CO($J$=2--1) and $^{13}$CO($J$=2--1) emission using the CSO and Mopra telescopes toward the DHN with angular resolutions of $\sim$33$''$, corresponding to $\sim$1.3 pc at the distance of the GC, 8 kpc.

\item Both molecular features, at $-35$ km s$^{-1}$ and $0$ km s$^{-1}$, are physically associated with the DHN. 
The 0 km s$^{-1}$ feature traces the infrared distribution of the DHN very well, indicating a clear association.
The association of the $-35$ km s$^{-1}$ feature is less clear compared with the 0 km s$^{-1}$ feature, but the complementary spatial distribution between the molecular gas and the DHN and the velocity variation along the DHN suggests a physical association with the DHN.

\item The molecular masses of the $-35$ km s$^{-1}$ and 0 km s$^{-1}$ features are estimated at $0.8\times10^4$ M$_\odot$ and $3.3\times10^4$ M$_\odot$, respectively.
The two molecular features also have high temperatures ($>$ 30 K), typical of GC clouds. 
The same heating mechanism that operates throughout the CMZ may therefore also be playing a dominant role in the DHN.
Stellar UV and cosmic rays are candidates for being important contributors to the gas heating.  

\end{enumerate}

\acknowledgments
This work was financially supported by a grant-in-aid for Scientific Research (KAKENHI, No. 21253003, No. 23403001, No. 22540250, No. 22244014, No. 23740149, No. 22740119, No. 24224005 and No. 23740149) from MEXT (the Ministry of Education, Culture, Sports, Science and Technology of Japan). This work was also financially supported by the Young Research Overseas Visits Program for Vitalizing Brain Circulation (R2211) and the Institutional Program for Young Researcher Overseas Visits (R29) by JSPS (Japan Society for the Promotion of Science) and by the grant-in-aid for Nagoya University Global COE Program, ``Quest for Fundamental Principles in the Universe: From Particles to the Solar System and the Cosmos'' from MEXT. MM acknowledges partial supported for this work by NASA/JPL subcontract 1316289 to UCLA.

\appendix
\section{Detailed distributions and conditions of the molecular features}
Here we present the detailed velocity structure of the molecular gas towards the DHN.
Figure \ref{lb2-1} shows the spatial distributions of the $-35$ and 0 km s$^{-1}$ features in CO($J$=2--1) emission.
The CO($J$=2-1) emission traces clumpy structures in the DHN compared with the CO($J$=1--0) emission in Figure \ref{lb}.
The $^{13}$CO emission traces only the dense components in the DHN, and has a negligible contribution from unrelated foreground emission.

Figure \ref{channel_all1} shows the velocity channel distributions of the $^{12}$CO($J$=1--0) emission with a velocity interval of 5.5 km s$^{-1}$.
The $-35$ and 0 km s$^{-1}$ features are seen around $-36$ -- $-30$ km s$^{-1}$ and $-2.7$ -- $2.8$ km s$^{-1}$, respectively.
The 10 km s$^{-1}$ feature, which seems to be unrelated to the DHN, is distributed around 8 -- 14 km s$^{-1}$.
The inter-velocity features seen in Figure \ref{allbv} occur in the velocity intervals at $-30$--$-19$ km s$^{-1}$ and $-19$--$-8$ km s$^{-1}$.

Figures \ref{channel_m35} and \ref{channel_0} show the detailed velocity channel distributions of the $-35$ km s$^{-1}$ feature and the 0 km s$^{-1}$ feature in the $^{12}$CO $J$=1--0 and $J$=2--1 transitions with a velocity separation 1.6 km s$^{-1}$.
In Figure \ref{channel_m35} the $^{12}$CO($J$=2--1) emission clearly shows that the two filaments at $\sim-34$ km s$^{-1}$ shows a complementary distribution with the DHN, and that the filament at $\sim-31$ km s$^{-1}$ coincides well with the DHN.
These velocity structures are clearly seen in Figure \ref{peakv}a.


Figures \ref{ratiochannel_m35} -- \ref{ratiochannel_10} show the velocity channel distributions of the  $^{12}$CO $J$=2--1/$J$=1--0 intensity ratio in the $-35$ km s$^{-1}$, 0 km s$^{-1}$ and 10 km s$^{-1}$ features.
All of these features have highly excited components that show ratios around 1.0.

\section{Absorption by foreground gas}
We make a model of a molecular cloud uniformly veiled by a layer of cool absorber in order to quantify the effect of the foreground gas on the intensities of CO lines of the DHN.
Here two sets of parameters are chosen for a hot molecular cloud and a cool molecular cloud; 
($T_{\rm k}$, $n$(H$_2$))~=~(30~K, 1000~cm$^{-3}$) and (10~K, 1000~cm$^{-3}$), respectively.
($T_{\rm k}$, $n$(H$_2$)) of the foreground absorber is assumed to be (10~K, 500~cm$^{-3}$).
Here we also assume that the foreground emission covers the velocity range of the DHN uniformly, having no frequency dependence, because no clear absorption feature is seen on the CO profiles (see Figure \ref{spec}). The velocity coverage of the foreground absorber is thus assumed to be the same or larger than that of the DHN features.
Distributions of the resulting brightness temperatures $T_{\rm obs}$ as a function of $X$(CO)/($dv/dr$) are shown in Figure \ref{lvgabs}, where $T_{\rm obs}$ is calculated with the following equation,
\begin{equation*}
T_{\rm obs} = J_\nu (T_{\rm cloud})(1-e^{-\tau_{\rm cloud}})e^{-\tau_{fg}} + J_\nu (T_{\rm fg}) (1-e^{-\tau_{\rm fg}}) - J_\nu (T_{\rm bg}) (1-e^{-(\tau_{\rm cloud}+\tau_{\rm fg})}),
\end{equation*}
where $\tau_{\rm cloud}$ and $\tau_{\rm fg}$ are the optical depths of the molecular cloud and foreground absorber, and $T_{\rm cloud}$, $T_{\rm fg}$ and $T_{\rm bg}$ = 2.7 K are the excitation temperatures of the molecular cloud, the foreground absorber and the cosmic microwave background, respectively.
($T_{\rm cloud}$, $\tau_{\rm cloud}$) and ($T_{\rm fg}$, $\tau_{\rm fg}$) are given with the LVG calculations.
Here we use $X$(CO)/($dv/dr$) of $2.5\times10^{-4}$ (km s$^{-1}$ pc$^{-1}$)$^{-1}$ same as in Section 3.3.
$J_\nu (T) = (h\nu/k_b)/(e^{h\nu/k_bT}-1)$, where $h$, $k_b$ and $\nu$ are the Planck constant, the Boltzmann constant and the rest frequency of the observed line, respectively.

\begin{figure}
\epsscale{.9}
\plotone{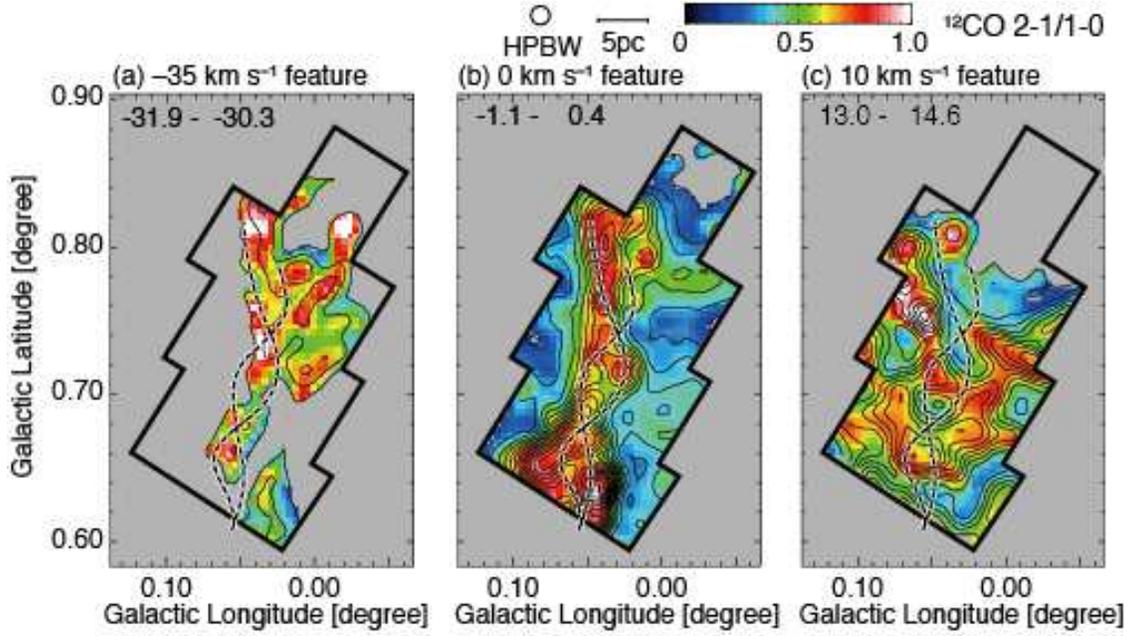}
\caption{Distributions of the $^{12}$CO($J$=2--1)/$^{12}$CO($J$=1--0) intensity ratio around $-31$ km s$^{-1}$, 0 km s$^{-1}$ and 14 km s$^{-1}$. 
Here only a part of the whole velocity range of each of the $-35$ km s$^{-1}$, 0 km s$^{-1}$ and 10 km s$^{-1}$ features is shown.
Contours show the integrated intensity of $^{12}$CO($J$=2--1) emission and are plotted at every 0.9 K km s$^{-1}$ from 0.7 K km s$^{-1}$. Maps of the individual velocity channel distributions within these three features are shown in Figures \ref{ratiochannel_m35}, \ref{ratiochannel_0} and \ref{ratiochannel_10} of the Appendix. \label{ratio_all}}
\end{figure}

\begin{figure}
\epsscale{.6}
\plotone{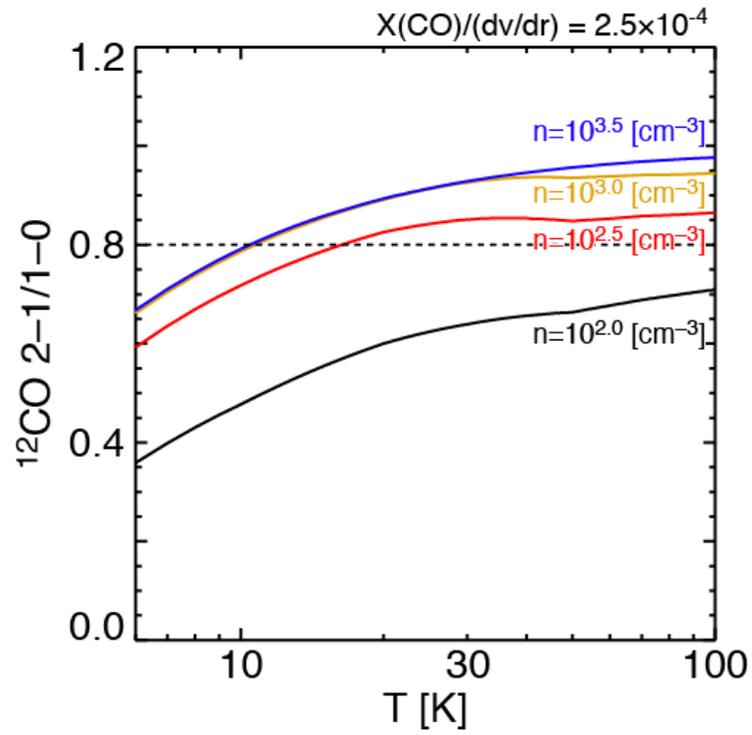}
\caption{Curves of the $^{12}$CO($J$=2--1)/$^{12}$CO($J$=1--0) ratio as a function of $T_{\rm k}$ and $n$(H$_2$) estimated using the LVG calculations. \label{lvg_plot}}
\end{figure}

\begin{figure}
\epsscale{1.}
\plotone{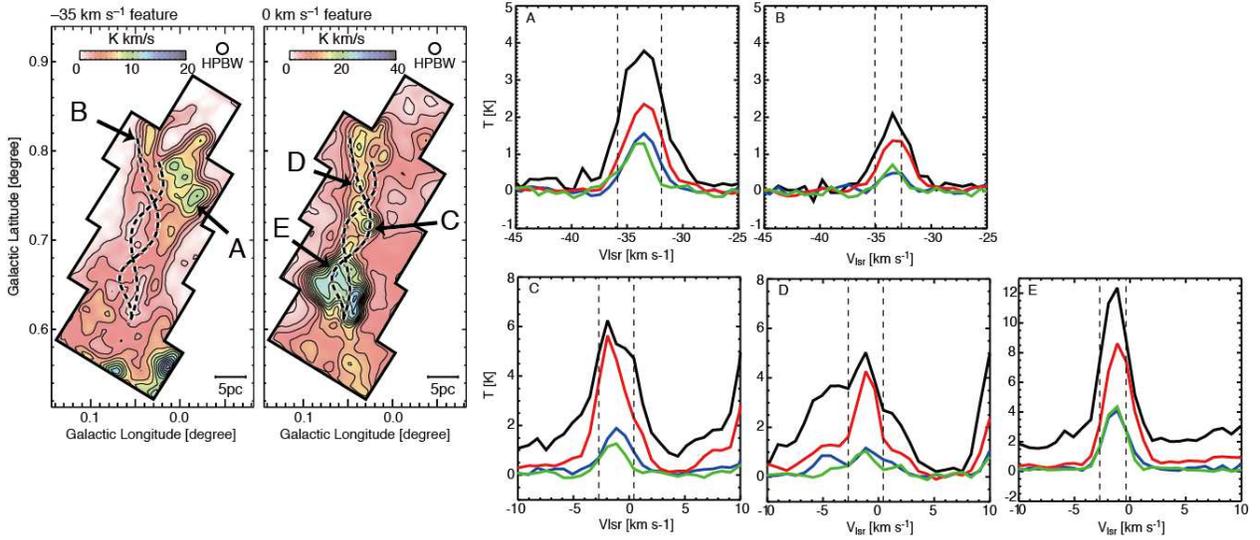}
\caption{(Left) The five regions A -- E used in the LVG analysis are depicted by arrows on the integrated intensity images of  $^{12}$CO($J$=2--1) emission. Contours are plotted at the same level as Figures \ref{lb+bv_0+m35}a and \ref{lb+bv_0+m35}c.
(right) Spectra toward the five peaks A -- E. Black, red, blue and green show profiles of $^{12}$CO($J$=1--0), $^{12}$CO($J$=2--1), $^{13}$CO($J$=1--0) and $^{13}$CO($J$=2--1) emission, respectively.\label{spec}}
\end{figure}

\begin{figure}
\epsscale{.9}
\plotone{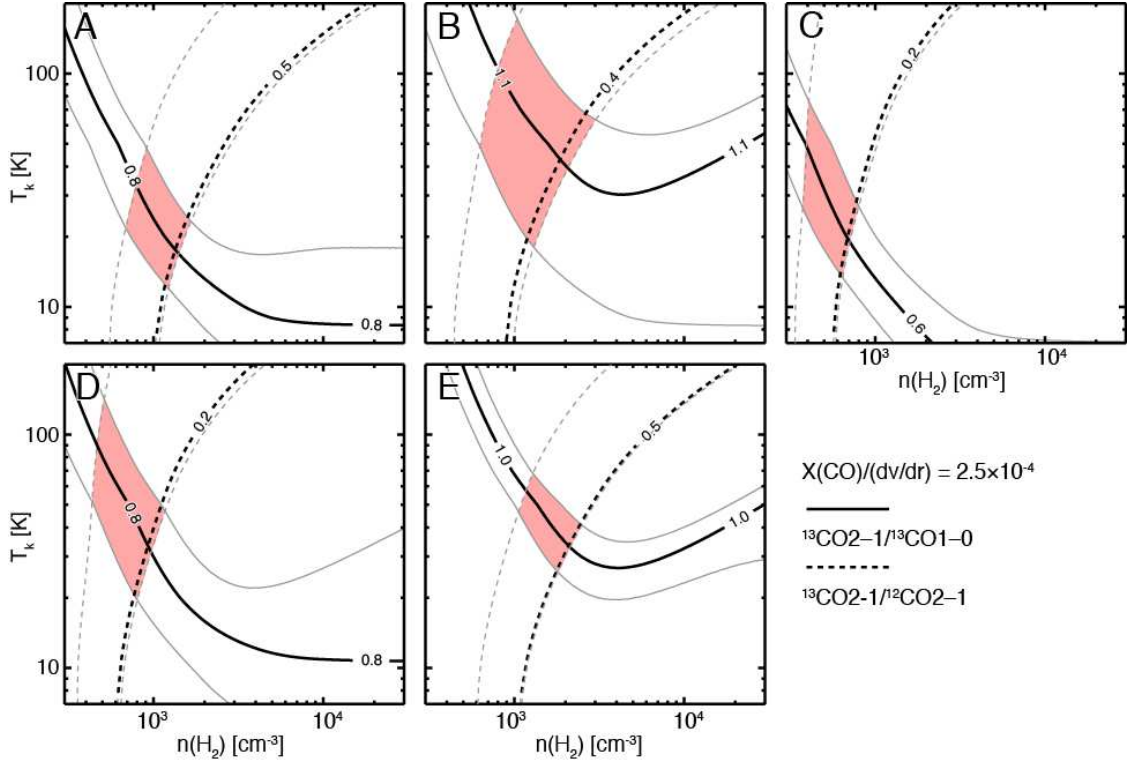}
\caption{Contour plots of the LVG results of the five peaks in Figure \ref{spec} are shown on the density-temperature plane. The $^{13}$CO($J$=2--1)/$^{13}$CO($J$=1--0) and $^{13}$CO($J$=2--1)/$^{12}$CO($J$=2--1) intensity ratios are depicted by solid lines and dotted lines, respectively. The filled area in each panel indicates the regions in which the two line ratios overlap within the errors. Details of the resulting parameters are summarized in Table \ref{lvglist}. \label{lvg}}
\end{figure}

In the case of the hot cloud in Figure \ref{lvgabs}(a),  the $^{12}$CO $J$=2--1/$J$=1--0 ratio increases to over 0.9 at $X$(CO)/($dv/dr$) $<$ $2\times10^{-6}$ (km s$^{-1}$ pc$^{-1}$)$^{-1}$.
The molecular features in the DHN show intensity ratios up to 1.1 as seen in Figure \ref{ratio_all}, consistent with the results of Figure \ref{lvgabs}(a).
In this $X$(CO)/($dv/dr$) range, which is depicted by the filled area in Figure \ref{lvgabs}, $T_{\rm obs}$ of $^{12}$CO($J$=1--0) is reduced by 30 -- 80 \%, and $^{12}$CO($J$=2--1) by 35 -- 85 \%. 
On the other hand, in the cool cloud case in Figure \ref{lvgabs}(b), the ratio varies over a range of 0.65 -- 0.85 and never beyond 0.9 at the $X$(CO)/($dv/dr$) range where the ratio is $>$0.9 in the hot cloud case in Figure \ref{lvgabs}(a).
This indicates that although the intensity ratio is strongly affected by the foreground gas, the CO spectra of the DHN with high ratios larger than 0.9 still indicate high excitation gas.

\clearpage

\clearpage

\begin{table}
\begin{center}
\caption{LVG results.\label{lvglist}}
\begin{tabular}{ccccc}
\tableline\tableline
 Region & $l$ & $b$ & $n$(H$_2$) (cm$^{-3}$) & $T_{\rm{k}}$ (K)  \\
\tableline

A 	& $-0\fdg02$	& 0\fdg74	& $1.3^{+0.3}_{-0.5}\times10^3$ 	& 17$^{+30}_{-5}$ 	 \\

B 	&$0\fdg04$	& 0\fdg81	& $1.9^{+1.1}_{-1.3}\times10^3$ 	& 43$^{+127}_{-25}$	 \\				
								
C	& $0\fdg03$	& 0\fdg72	& $6.9^{+1.1}_{-3.2}\times10^2$ 	& 20$^{+58}_{-7}$	 \\

D	& $0\fdg04$	& 0\fdg77	& $9.3^{+1.7}_{-5.0}\times10^2$ 	& 34$^{+96}_{-15}$		 \\
				
E 	& 0\fdg04		& 0\fdg63	& $2.1^{+0.3}_{-1.1}\times10^3$ 	& 33$^{+36}_{-10}$	 \\

\tableline
\end{tabular}
\tablecomments{Column (4): Number density of H$_2$. (5): Kinetic temperature. }
\end{center}
\end{table}

\clearpage

\begin{figure}
\epsscale{.90}
\plotone{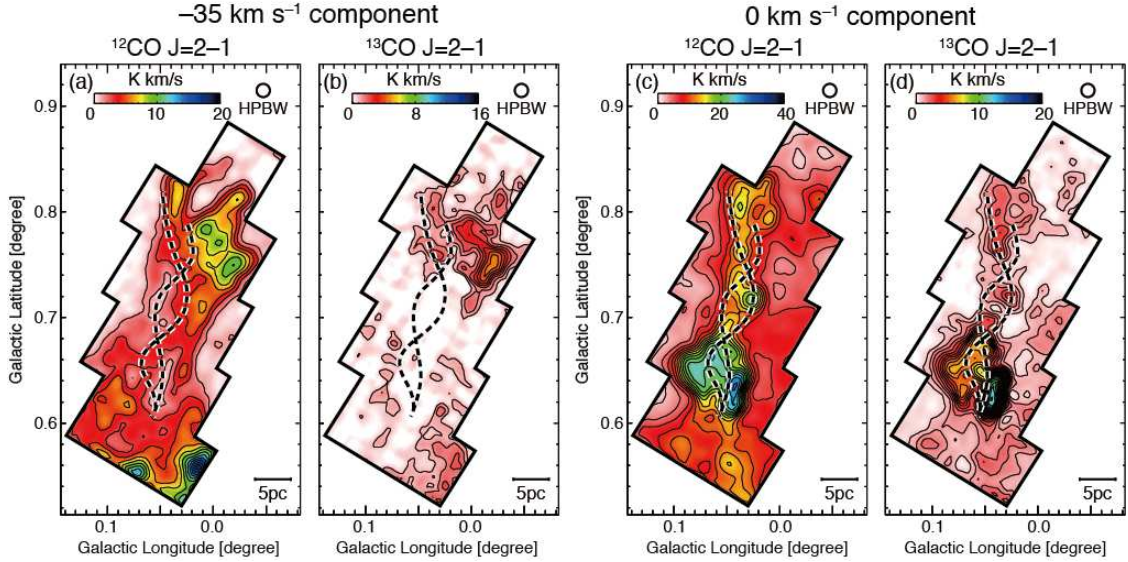}
\caption{Distributions of the two CO features in the CO($J$=2--1) emission. The $-35$ km s$^{-1}$ feature are shown in panels (a) and (b), and the 0 km s$^{-1}$ feature is in panels (c) and (d). Panels (a) and (c) show the $^{12}$CO($J$=2--1), and panels (b) and (d) is $^{13}$CO($J$=2--1). The contour levels of each panels are as follows; (a) minimum: 2 K km s$^{-1}$, step: 1.5  K km s$^{-1}$. (b) minimum: 0.8 K km s$^{-1}$, step: 0.6 K km s$^{-1}$. (c) minimum: 2 K km s$^{-1}$, step: 2 K km s$^{-1}$. (d) minimum: 0.8 K km s$^{-1}$, step: 0.6 K km s$^{-1}$.
\label{lb2-1}}
\end{figure}

\begin{figure}
\epsscale{.80}
\plotone{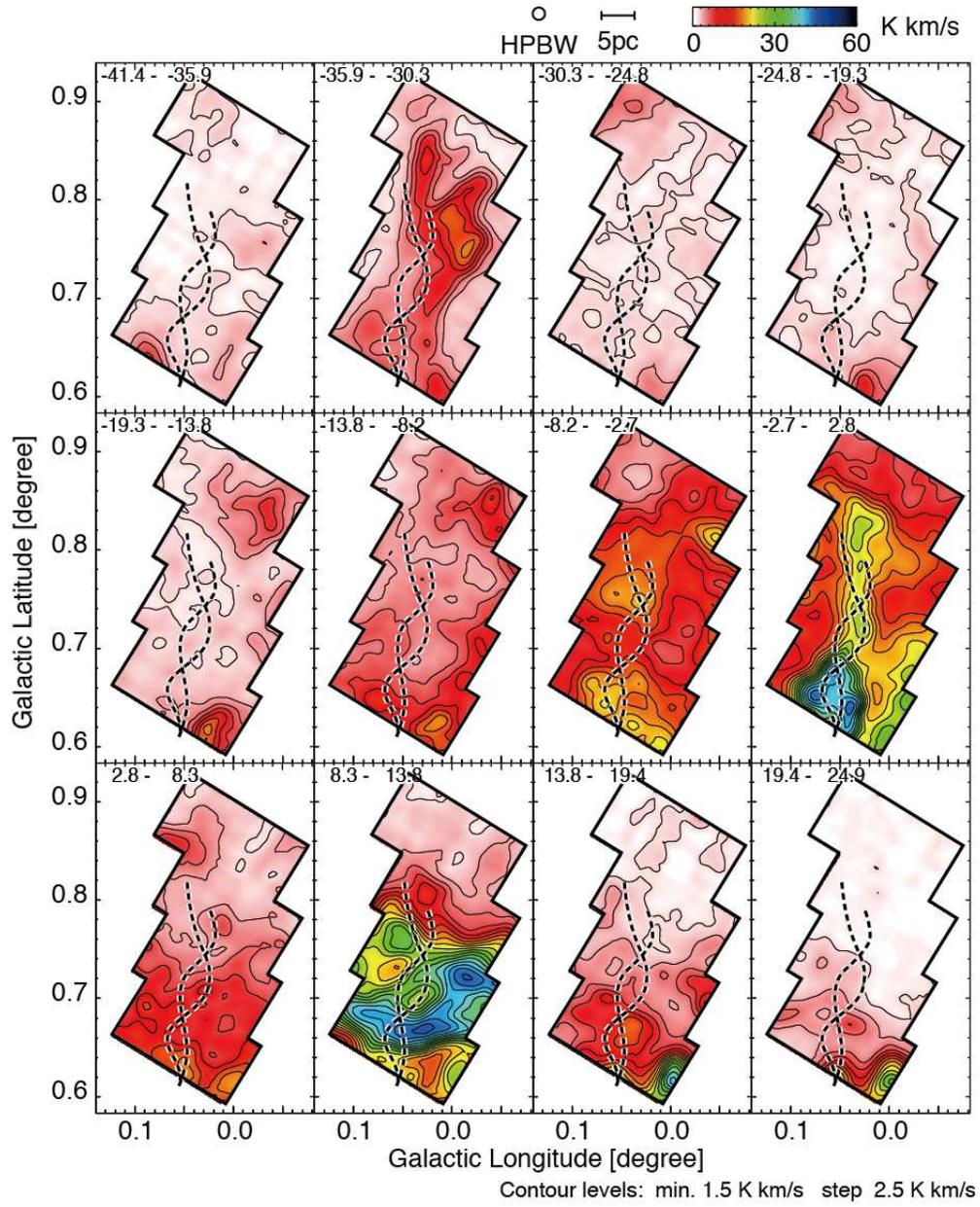}
\caption{Velocity channel distributions of $^{12}$CO($J$=1--0) toward the DHN. \label{channel_all1}}
\end{figure}

\begin{figure}
\epsscale{.80}
\plotone{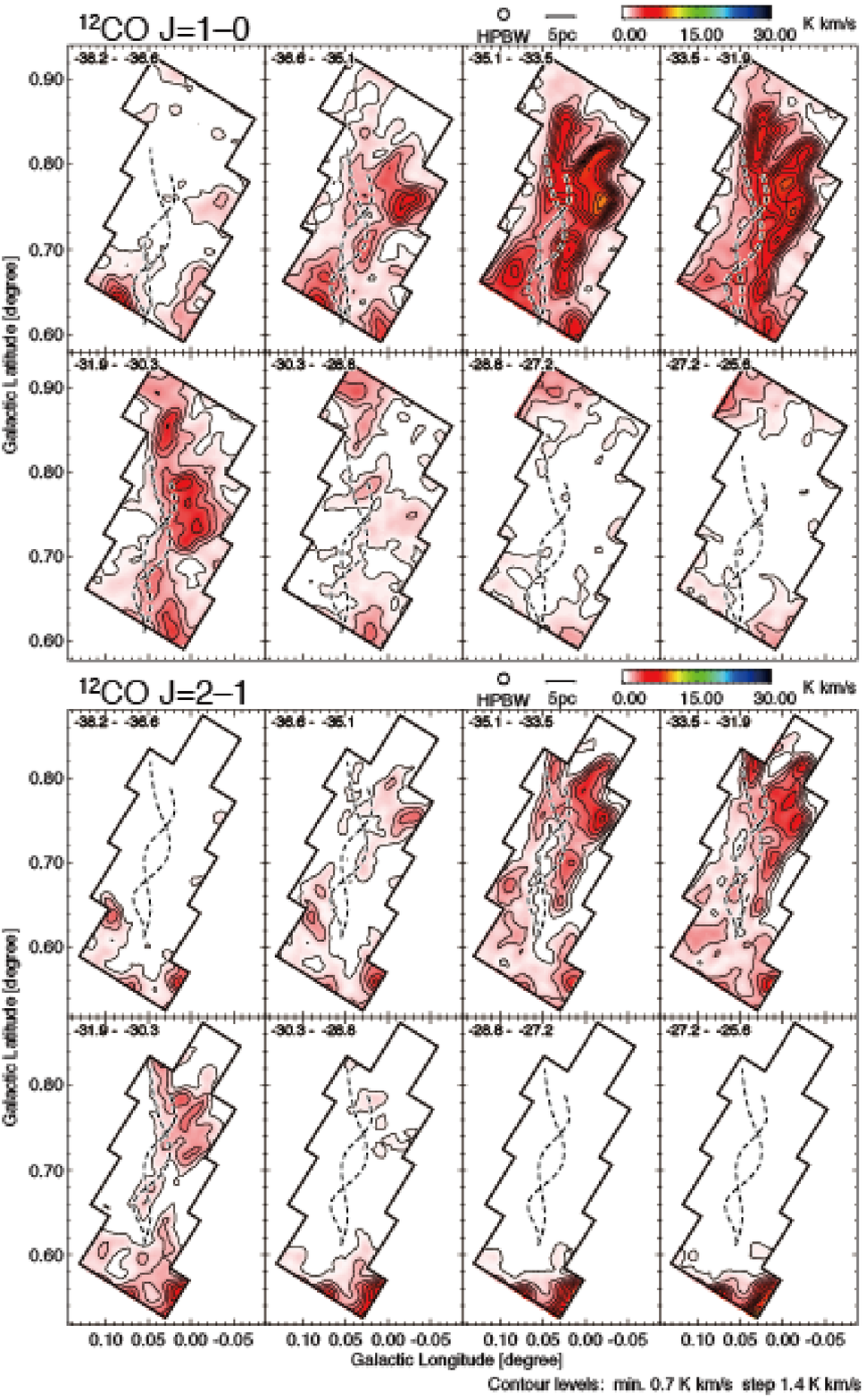}
\caption{Detailed velocity channel distributions of $^{12}$CO($J$=1--0) and $^{12}$CO($J$=2--1) toward the $-35$ km s$^{-1}$ feature. Upper eight panels show $^{12}$CO($J$=1--0) and lower eight panels show $^{12}$CO($J$=2--1). \label{channel_m35}}
\end{figure}

\begin{figure}
\epsscale{.80}
\plotone{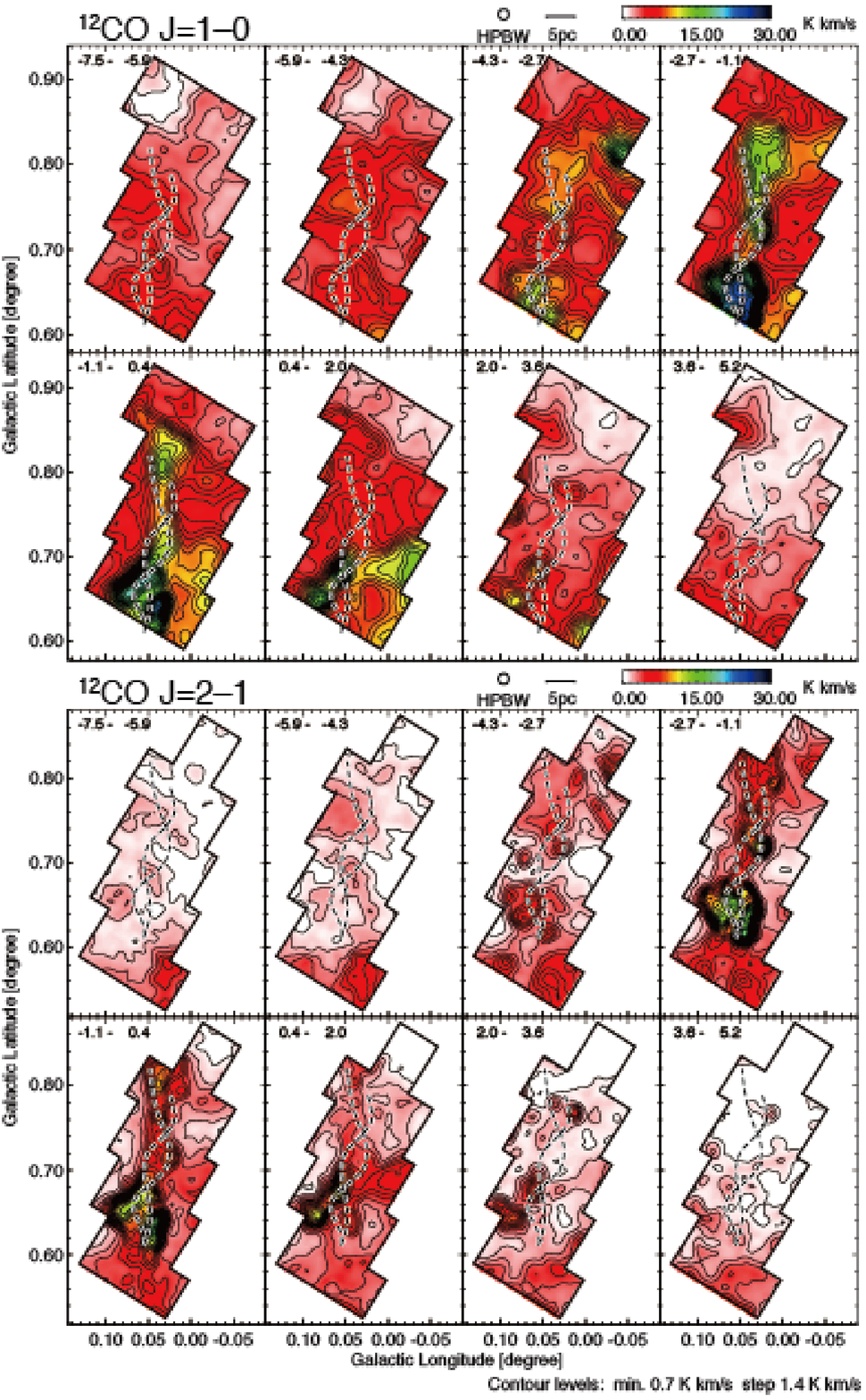}
\caption{Detailed velocity channel distributions of $^{12}$CO($J$=1--0) and $^{12}$CO($J$=2--1) toward the 0 km s$^{-1}$ feature. Upper eight panels show $^{12}$CO($J$=1--0) and lower eight panels show $^{12}$CO($J$=2--1). \label{channel_0}}
\end{figure}

\begin{figure}
\epsscale{.9}
\plotone{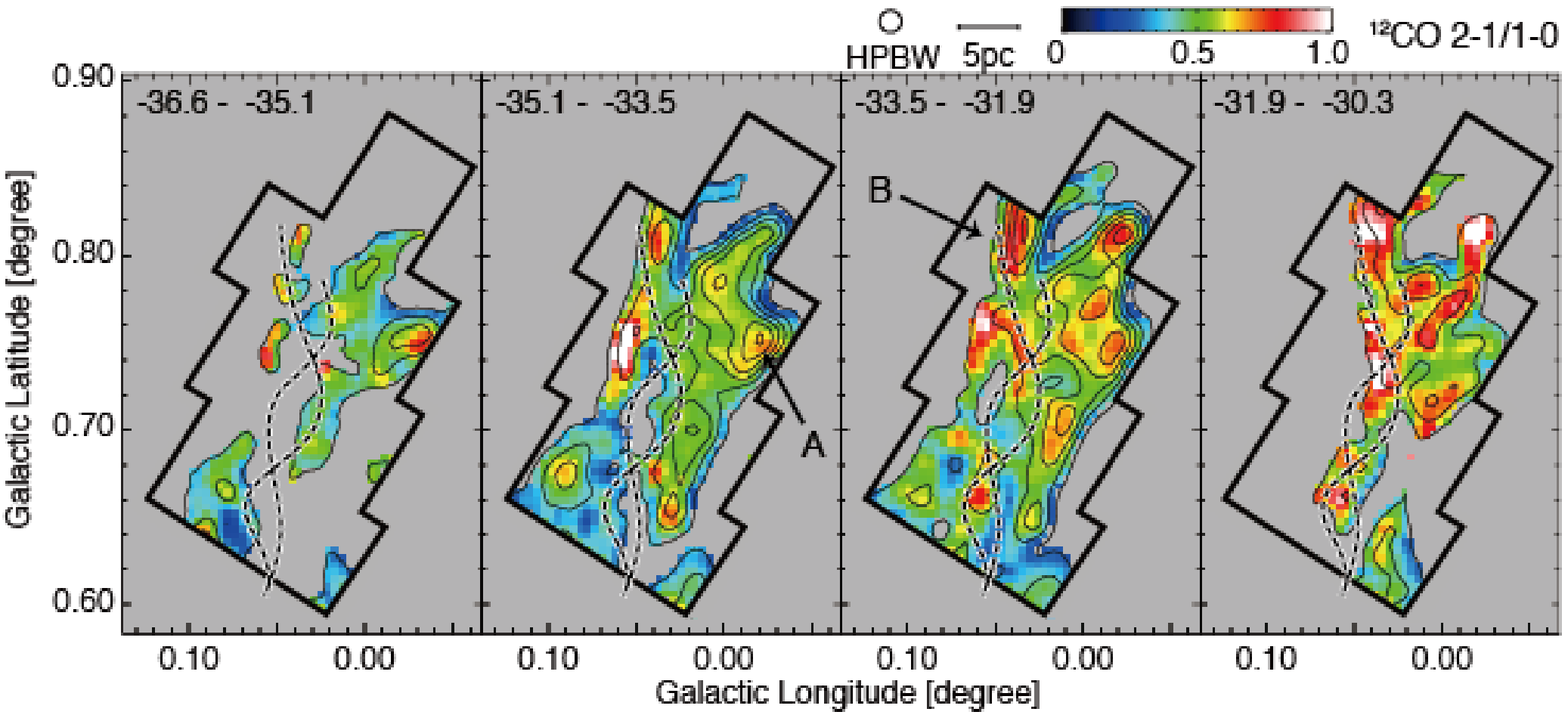}
\caption{Velocity channel distributions of $^{12}$CO($J$=2--1)/$^{12}$CO($J$=1--0) ratios toward the $-35$ km s$^{-1}$ feature. Contours show the $^{12}$CO($J$=2--1) emission and are plotted at every 0.9 K km s$^{-1}$ from 0.7 K km s$^{-1}$.\label{ratiochannel_m35}}
\end{figure}

\begin{figure}
\epsscale{.9}
\plotone{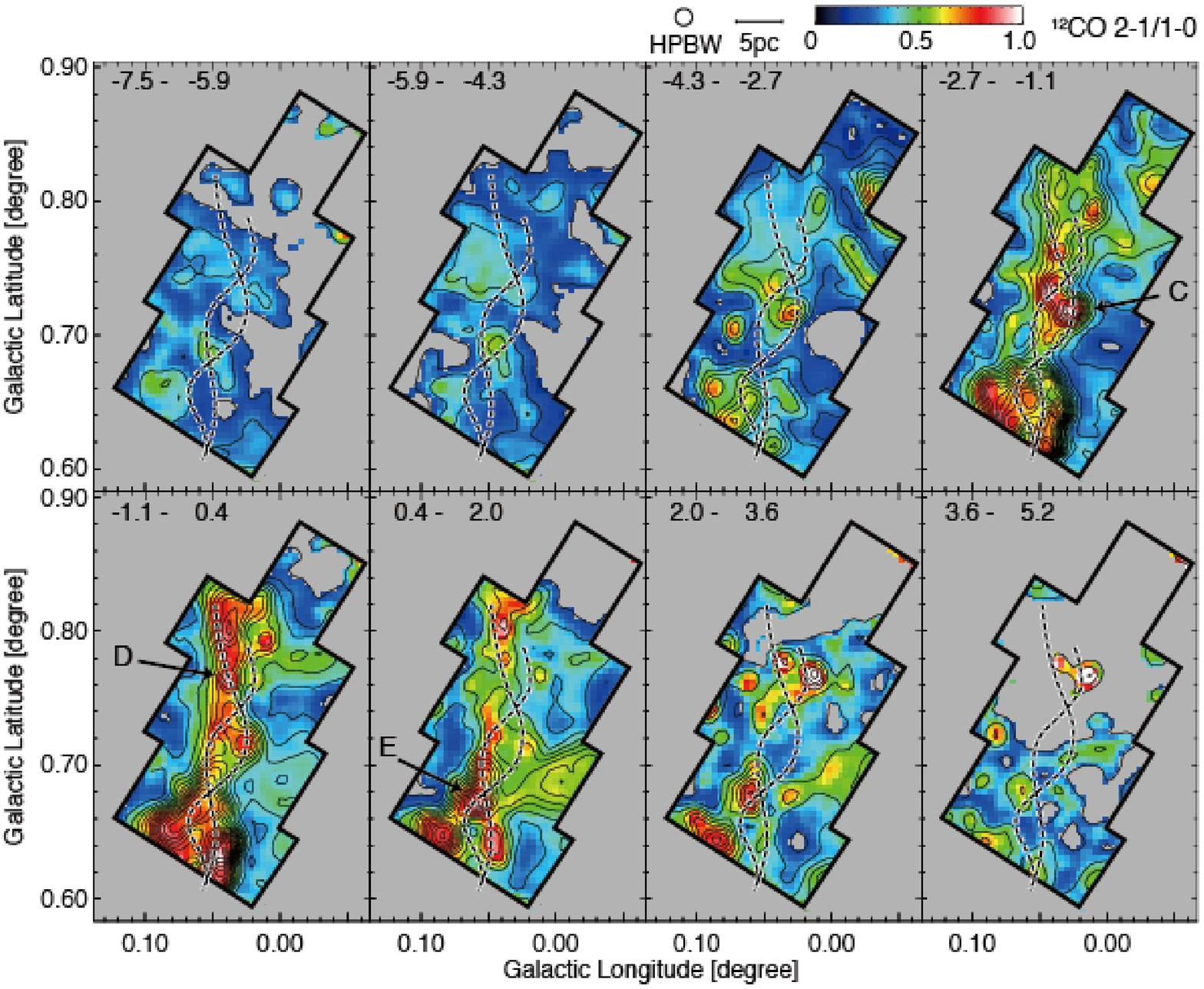}
\caption{Velocity channel distributions of $^{12}$CO($J$=2--1)/$^{12}$CO($J$=1--0) ratios toward the $0$ km s$^{-1}$ feature. Contours show the $^{12}$CO($J$=2--1) emission and are plotted at every 0.9 K km s$^{-1}$ from 0.7 K km s$^{-1}$.\label{ratiochannel_0}}
\end{figure}

\begin{figure}
\epsscale{.9}
\plotone{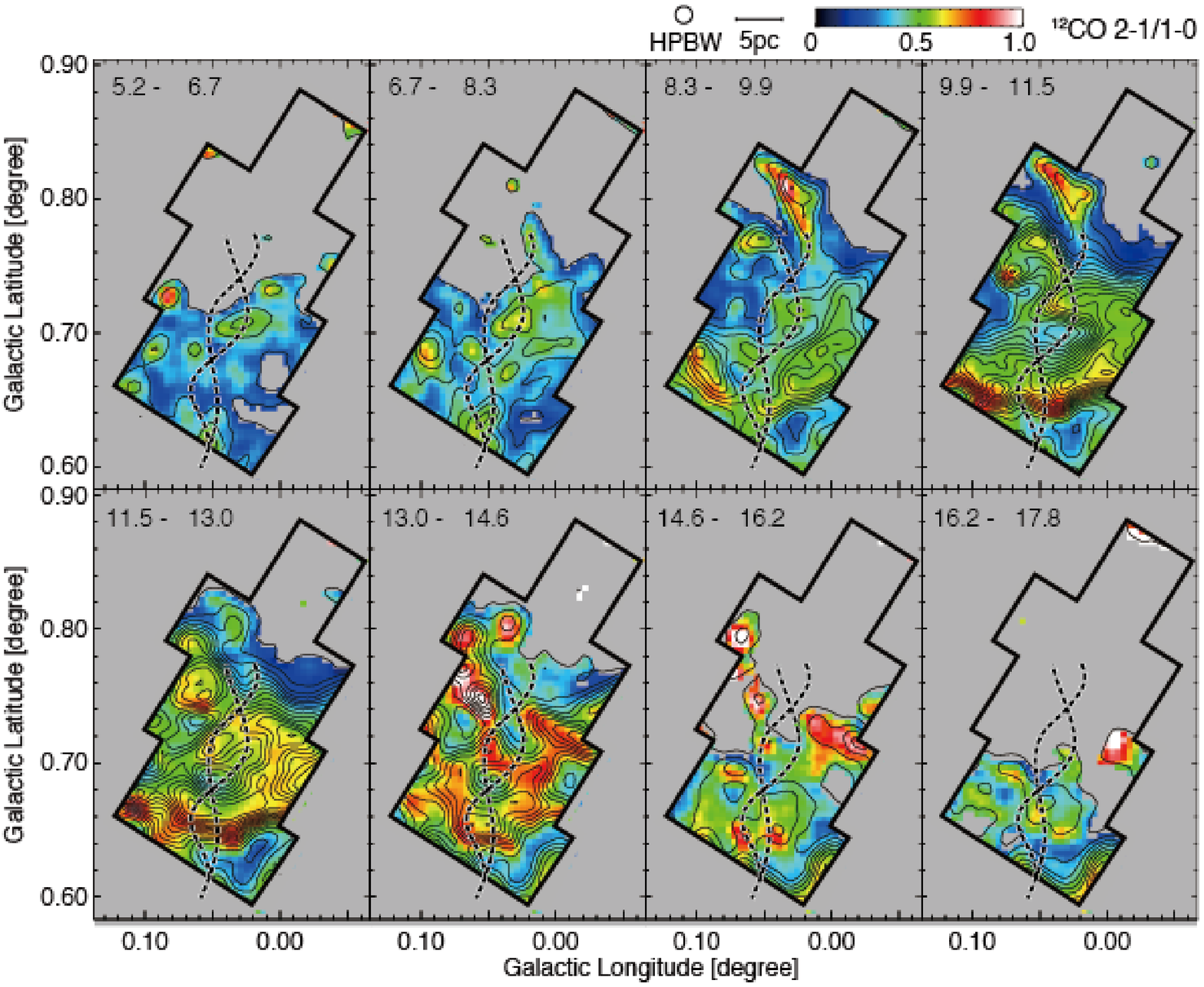}
\caption{Velocity channel distributions of $^{12}$CO($J$=2--1)/$^{12}$CO($J$=1--0) ratios toward the $10$ km s$^{-1}$ feature. Contours show the $^{12}$CO($J$=2--1) emission and are plotted at every 0.9 K km s$^{-1}$ from 0.7 K km s$^{-1}$.\label{ratiochannel_10}}
\end{figure}

\begin{figure}
\epsscale{0.9}
\plotone{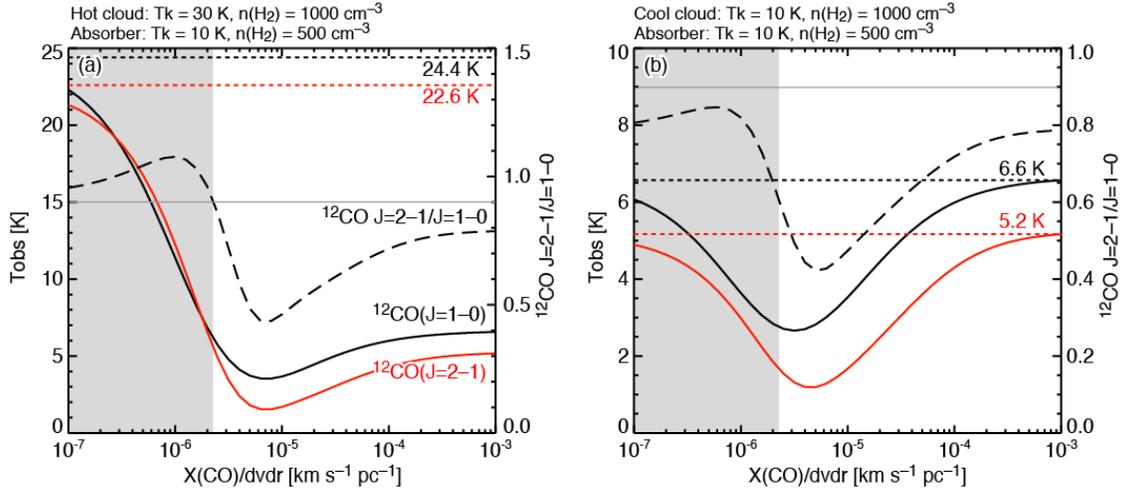}
\caption{Curves of the $^{12}$CO($J$=1--0) and $^{12}$CO($J$=2--1) brightness temperatures derived with the LVG analysis as a function of  $X$(CO)/($dv/dr$). Here two sets of the parameters of molecular cloud are shown; (a) $T_{\rm k}$~=~50~K and $n$(H$_2$)~=~10$^3$~cm$^{-3}$, and (b) $T_{\rm k}$~=~10~K and $n$(H$_2$)~=~10$^3$~cm$^{-3}$. Solid lines show the temperatures absorbed by the foreground cool absorber, and dashed line shows the ratios between the two transitions. Dotted lines show the original temperatures not affected by the foreground gas. The filled areas show the region with ratios larger than 1.0. \label{lvgabs}}
\end{figure}

\end{document}